\documentclass{optica-article}

\journal{opticajournal} 

\articletype{Research Article}

\usepackage{lineno}

\begin{document}

\title{Quantum Photonic Gates with Two-Dimensional Random Walkers}

\author{S. Ali Hassani Gangaraj,\authormark{1,*} and Dan T Nguyen,\authormark{1,$\dagger$}}

\address{\authormark{1}Optical Physics Division, Corning Research and Development, Corning, New York, 14831, USA}

\email{\authormark{*}hassanigs@corning.com} 
\email{\authormark{$\dagger$}nguyendt2@corning.com} 


\begin{abstract*} 
Quantum gates are crucial for processing quantum information, but implementing them in a photonic platform poses unique challenges due to the peculiar way photons propagate and interfere. Here, we examine quantum photonic gates that utilize continuous time two-dimensional random walking photons. These gates can be implemented using the inverse design method, where photons randomly walk in a two-dimensional silicon host medium embedded with silicon dioxide scatterers. We propose a C-NOT gate as a multiqubit gate and an X-gate as a single qubit gate. In addition, we provide studying the non-trivial spatial correlations of random walking photons by utilizing the quantum correlation function. The results demonstrate high-fidelity probabilistic quantum gates. Further work is required to address error-correction. This work advances the practical implementation photonic elements in linear optics quantum computation schemes.
\end{abstract*}

\section{Introduction}
Quantum information processing (QIP) represents a broad research area with numerous applications in quantum computation, communication, and information. In general, these applications would require operations by a number of gates known as universal quantum gates \cite{D1,D2,D3,D4}. So far, these quantum gates have been successfully implemented with ion-trapped, ultra-cold atoms and superconducting technologies \cite{D5,D6,D7,D8,D9,D10} which are the main reasons for impressive progresses of quantum computing based on those techniques. However, the quantum gates that are based on these technologies must operate at extremely low temperatures, often at tens of millikelvin (mK) and in extremely high vacuum conditions. This working principle would pose a very challenging problem for the future uses of these quantum devices. Meanwhile, quantum gates and quantum computing protocols based on photons in linear optics operating at room temperature were also proposed. In 2001, Knill, Lafflame and Milburn had proposed an efficient quantum computation that is possible with linear optics (KLM protocol) \cite{D11}. 

The early promising proposals of linear optics quantum computing (LOQC) have inspired researchers to explore quantum computing using photonics technology. Meanwhile, photonics technology has become mature with many photonics systems, including micro- and nano-scale ones, operate ideally at room-temperature without highly vacuum conditions. That is one of the advantages of photonics as compared to other technologies. However, despite great efforts, the successful results of quantum photonics computing including quantum photonics gates have been very limited \cite{D13,D14,D15}, and many obstacles have not been successfully resolved. There are a number of reasons for this. First and foremost, although the KLM protocol has shown an efficient LOQC is possible but the gate implementations with linear optics provides only probabilistic outcomes, and resource for high probability operations are extremely demanding. Second, the LOQC schemes are too difficult to be implemented in integrated photonics platforms, especially for large-scale computation. Furthermore, the LOQC schemes are not completely linear: the operations still require so-called hidden measurement-induced nonlinearity \cite{D11}. In other words, nonlinearities are still unavoidable in the linear optics KLM protocols, and the scheme has been unsuccessfully developed so far, after nearly 25 years. Naturally, ideas of photonics quantum computing operations based on nonlinear effects have been proposed. For examples, the authors of \cite{D16} proposed to use Kerr nonlinearities in bulky devices and are therefore not suitable for real applications. On the other hand, some other proposals suggest that the nonlinear optics of a single photon may yet become an important direction for quantum gates \cite{D17}. However, so far there have been no successful implementation of quantum gates based on that scheme. 

It is clear that quantum computing with linear optics if successfully implemented would be a huge advantage as compared with the ones using nonlinear optics. First, although optical nonlinearity has been used in the early all-optical switches which are essential in many logic gates, it is very weak and noisy at single or few photon regime. That’s because photons do not interact with each other in free space but only in the presence of electrons through a mediated material, and the strength of optically nonlinear interaction is much weaker than that of electrons. Therefore, nonlinear photon-matter interaction requires considerable power and it is dissipative. Second, even if nonlinearity for switching is achievable for single or small numbers of gates, it is still not easy to control a large number of gates in large-scale quantum computing in which thousands or much higher numbers of gates are used. In such situations, designs of simple but effective ways to control are very much needed for a breakthrough. The situation of photonic quantum computing can be summarized as follows: for LOQCs which are based on linear optics, the quantum gates operate probabilistically, and high probability outcomes are very resource demanding. Moreover, the hidden measurement-induced nonlinearity in LOQC is also too complicated for realization. On the other hand, nonlinear optics quantum computing (NLOQC) that relies on quantum gates with deterministic operation would require a breakthrough of nonlinear optics materials and/or designs that enable switching on-chip devices effectively for large scales quantum computing systems.

In 2018, a new approach of designing quantum logic gates has been proposed by Y. Lahini et al., \cite{D18} based on quantum walk. Quantum Walks are unitary processes that describes the evolution of quantum particles with wave-particle duality on a potential lattice \cite{Aharonov,Farhi}. For the last 20 years, there has been significant of efforts in the investigation of QWs in photonics lattices including arrays of beam splitters \cite{D19}, arrays of waveguides \cite{D20,D21}, multicore fiber (MCF) \cite{D22}. Photons, the quanta of electromagnetic field are considered as excellent “walkers” as their generation and manipulation are highly controllable. From the very fundamental research in early 2000s, QWs have become one the key operations in a number of quantum protocols of quantum computing such as modelling for exponential speedup in quantum algorithms \cite{D23,D24,D25,D26}, to implement universal quantum gates for quantum computers \cite{D27,D28} and quantum simulations \cite{D30}. Especially, the QW-based Boson sampling \cite{D31,D32} are not only the demonstration of quantum computational advantage but also could have important impacts on material science \cite{D33,D34}. 

The design approach based on photon quantum walk in \cite{D18} is based on a simple one-dimensional array of potential wells using minimal resources: one quantum walker and a small number of lattice sites per qubit. The authors demonstrate how to control the lattice potential of quantum walk (QW) can impose certain spatial correlations between walkers, and the results of QW’s distributions can be used as outcomes of quantum photonics logic gates (QPLG). For non-interacting bosons (such as photons in waveguide lattices) they find that this approach yields high-fidelity probabilistic logic gates with a similar success rate to those found previously - but with a much simpler physical design. For interacting bosonic quantum walkers (e.g., ultra-cold atoms in optical potentials) they find that a complete set of high-fidelity QPLG can be realized using a linear array of potential wells with nearest-neighbor coupling and only two sites per qubit, demonstrating the universality of this architecture for quantum computation. One of the most important features of the non-interacting bosons quantum gates is the simple concept of utilizing quantum walks in arrays of waveguides which can be implemented near-perfectly within current technology. In their approach, the dual-rail qubits with state $ \left | 0 \right > $ and $ \left | 1 \right > $ of a qubit are defined as photon states occupied the left and right waveguides, respectively, of the waveguide lattice. It can be seen that the dual-rail qubits work very naturally as quantum walkers in the photonics lattice, and the operations of quantum gates can be realized just by the results of quantum walk process of the qubits on the platform of waveguide arrays. 

In this work, we build upon the concept of QPLG operating on dual-rail qubits walking on arrays of waveguides, as previously explored in \cite{D18}. However, instead of relying on fixed and simple arrays of waveguides, we delve into more intricate structures by utilizing the inverse design method \cite{Hassani_JLT, Zongfu, Zongfu2, Zongfu3} to enhance the probability outcomes of the logic gates. The concept of random quantum walks in this work, therefore is more general than either the conventional continuous-time quantum walk in arrays of waveguides \cite{D18,D20,D24} or discrete-time quantum walk in beam splitters \cite{D19}. Furthermore, we present a quantum description of QPLGs which is essential for quantum computing because it captures the inherently probabilistic quantum mechanical phenomena that classical descriptions cannot. This allows for more complex and efficient computation, leveraging quantum parallelism and interference effects that are fundamental to quantum algorithms and protocols. Unlike the classical description we provided in \cite{Hassani_JLT}, the quantum correlation function description provided here ensures accurate optimization and development of gates for practical quantum computing systems.

To initiate our discussion, we examine the case of multiple random walkers in a two-dimensional lattice. Within the quantum formalism, we present the corresponding quantum correlation matrix and fidelity. To assess the efficiency of this gate, we employ a two-fold coincidence detection scheme. The results demonstrate that the proposed C-NOT gate achieves a success probability of approximately 0.46, compared to 1/9 as seen in previous studies \cite{D18, Ralph}, with a fidelity approaching 0.95. Next, we simplify the scenario to a single random walker in a two-dimensional lattice. In this case, we propose a photonic X-gate to realize the desired outcome and evaluate its efficiency using a one-fold coincidence detection scheme. Our findings illustrate that for the proposed X-gate, both the success probability and fidelity approach 1.


\section{Two-Dimensional Quantum Random Walk for Computing}

There are several challenges in using random walk for quantum gates. These challenges encompass the physical demonstration of (I) the random walker, (II) the quantum bit or \emph{qubit} which is the basic element of interest for quantum gates, and (III) a platform through which multiple random walkers can propagate simultaneously.

To address these challenges, we propose using photons as the quantum walkers. We define the qubits using a dual-rail encoding, where a qubit is implemented by a single photon in a pair of coupled photonic waveguides. The states of the qubit, represented as $ \left|0 \right> $ and $ \left| 1 \right> $, correspond to the photon being in the top or bottom waveguide respectively (see Fig. \ref{fig1}(a)). Each quantum walker provides a qubit for the computation. More importantly, we propose a new photonics platforms for quantum random walks to realize logic operation. This environment consists of a host medium and a group of embedded scatterers through which the photon randomly walks.

We begin with a C-NOT gate, operating based on two-dimensional random walkers. We then move on to the case of a single qubit gate, such as the X-gate, which operates based on a single two-dimensional random walker.


\subsection{Multiple Random Walkers: C-NOT Gate}

\begin{figure}[ht!]
	\centering\includegraphics[width=12cm]{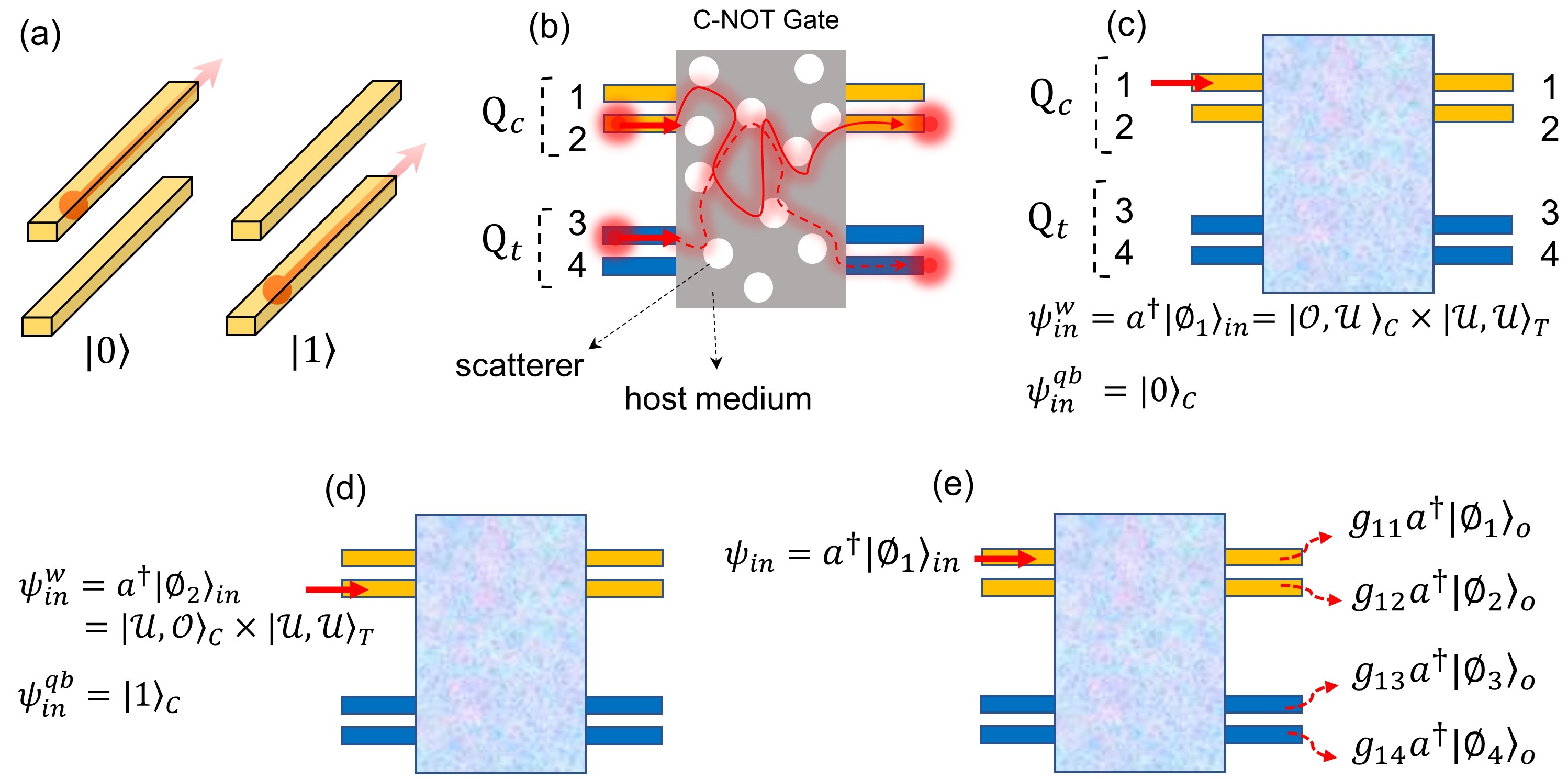}
	\caption{(a)- Dual-rail demonstration of a quantum photonic bit where the states $ \left | 0 \right > $ and $ \left | 1 \right > $ are defined by the photon being in the top or bottom waveguide. (b)- Physical representation of a quantum photonic C-NOT gate. (c) and (d)- Represent two different possible positions of an input photon and illustrates the corresponding waveguide-occupied state $ \psi^{w} $, and qubit-state $ \psi^{qb} $, presentations. (e) Illustrating a general state evolution of a photon created in the first waveguide of a C-NOT gate. The state $ \left | \varnothing_{i}\right>_{in} $ and $ \left | \varnothing_{i}\right>_{o} $ represents the i$th$ input and output waveguide vacuum state, respectively.}\label{fig1}
\end{figure}

A C-NOT gate, which stands for Controlled NOT gate, is a fundamental building block in either quantum or classical
computing. It is a two-qubit gate that performs a NOT operation (flipping the state from $\left | 0 \right> $ to $\left | 1 \right> $ or vice versa) on the target qubit, $Q_t$ only if the control qubit, $Q_c$ is in the state $\left | 1 \right> $. The proposed C-NOT gate here, consists of two pairs of single mode wavguide as the input and output control $( Q_c )$ and target $ (Q_t) $ qubits, see Fig. \ref{fig1}(b). As mentioned before, the definition of qubit is based on the 
dual-rail encoding for both target and control qubits. The qubits states are defined based on the photon's site in each waveguide pair, and the allowed states for each qubit are $ \left | 0 \right > $ and $ \left | 1 \right > $ when a photon occupies the top or bottom waveguide in a waveguide pair. For instance in Fig. \ref{fig1}(b) input waveguides 1 and 2 form the input control qubit with a photon in the bottom waveguide while the waveguides 3 and 4 form the input target qubit with a photon occupying the top waveguide. This indicates the input state as $ \psi^{qb}_{in} =  \left |  Q_c \right > \left| Q_t \right > =   \left |  1 \right >_C \left| 0 \right >_T $.

In addition to the two input waveguide pairs, two output pairs form the output control and target qubits. The two injected input photons undergo random walk within the gate and appear in the output waveguides. For example, in Fig. \ref{fig1}(b), the two input photons from waveguides 2 and 3 appear in waveguides 2 and 4, indicating the output state as $  \left |  1 \right >_C \left| 1 \right >_T $ where compared to the input state, the target qubit state is flipped. The operational principle of the gate is based on in the photons' two-dimensional random walk within the gate's main body, composed of a two-dimensional host and a group of scatterers (depicted as the gray area and distributed white particles in Fig. \ref{fig1}(b), respectively). Note that, in Fig. \ref{fig1} two different presentations have been used for the same photon states, the \emph{qubit-state} presentation $ \psi^{qb} $, and \emph{waveguide-occupied state} presentation $\psi^w$. The details of the presentations will be discussed in the following.    

So far, the input and output states of the logic gates has been presented in the qubit-state presentation. However, it can be presented in the waveguide-occupied state presentation as well. The possible states corresponding to waveguide-occupied presentation is classified as "occupied" $ \left| \mathcal{O} \right> $ or "unoccupied" $ \left| \mathcal{U} \right> $ depending on whether a photon is "present" or "absent" at the respective waveguide site. For example, in a system of four waveguides shown in Fig. \ref{fig1}, the state $  \psi^w_{in}  = a^{\dagger} \left| \varnothing_{i} \right>_{in}; ~ 1, ~ 2, ~ 3, ~ 4 $ indicates the creation of a photon in i$th$ input waveguide where $\left| \varnothing_{i} \right>_{in}$ represents its vacuum state. For instance in Fig. \ref{fig1}(c) one photon is created in the first waveguide. Since there is only one photon in the top waveguide of the pair defining the control qubit,this state can be presented in qubit-state as:
\begin{equation}\label{C-NOT1}
	\psi^{qb}_{in} =  \left | 0 \right>_C
\end{equation}
however in the waveguide-occupied state, it is demonstrated as: 
\begin{equation}\label{waveguide-occupied}
	\psi^w_{in} = a^{\dagger} \left| \varnothing_{1} \right>_{in}  \rightarrow \left| \mathcal{O}, \mathcal{U} \right>_C \left| \mathcal{U} , \mathcal{U}  \right>_T
\end{equation}
where the single entry state in Eq. \ref{C-NOT1}, $ \left | \cdots \right >_C $ denotes the state of the control qubit with possible entries of $0$ and $1$. But the double-entry state in Eq. \ref{waveguide-occupied}, $ \left| \dots, \dots \right>_C \left| \dots, \dots \right>_T $ denotes the presence or absence of photon in waveguides corresponding to the control (waveguides 1 and 2) and target (waveguides 3 and 4) qubits, respectively. 

As another example, Fig. \ref{fig1}(d) illustrates the creation of a photon in the second waveguide. Therefore the waveguide state can be expressed as $\psi^w_{in} = a^{\dagger} \left| \varnothing_{2} \right>_{in} $ meaning one photon is created in the second waveguide, or
\begin{equation}\label{C-NOT2}
	\psi^w_{in} = \left| \mathcal{U}  , \mathcal{O} \right>_C \left| \mathcal{U}, \mathcal{U}  \right>_T \rightarrow \psi^{qb}_{in} = \left | 1 \right>_C
\end{equation}
where $ \left|  \mathcal{U}, \mathcal{O} \right>_C $ denotes that the top waveguide of the control qubit waveguide pair is empty, and there exists one photon in the bottom waveguide, indicating state $ \left | 1 \right >_C $ for the control qubit. On the other hand $\left| \mathcal{U} , \mathcal{U}  \right>_T$ means there is no photon in waveguie pair defining target qubit state. Therefore the expression in Eq. \ref{C-NOT2} means that the input control qubit is in the state $ \psi^{qb}_{in} =  \left | 1 \right>_C $.

Based on the above-mentioned definition, when two photons are created in the input waveguides, see Fig. \ref{fig1}(b), one can write the in waveguide and qubit-states as follow

\begin{align}
	& \psi^w_{in} = a^{\dagger} \left| \varnothing_{2} \right>_{in} \times  a^{\dagger} \left| \varnothing_{3,in} \right>_{in}  \nonumber \\&
	\psi^w_{in} =  \left| \mathcal{U}  , \mathcal{O} \right>_C \left| \mathcal{O}, \mathcal{U}  \right>_T = \left | 1 \right>_C \left | 0 \right>_T
\end{align}
where in terms of waveguide states, two photons are simultaneously created in 2nd and 3rd waveguides $	\psi^w_{in} =  \left| \mathcal{U}  , \mathcal{O} \right>_C \left| \mathcal{O}, \mathcal{U}  \right>_T $, therefore the input control and target qubits are prepared in the state $ \psi^w_{qb} = \left | 1 \right>_C \left | 0 \right>_C $. Next we use this formalism to calculate the quantum correlation matrix.


\subsubsection{Correlation Functions: Two-fold Probability}

The created photon in the input waveguide of the gate shown in Fig. \ref{fig1}, randomly walks through the host and scattereres and will appear in the output ports. Therefore its evolution can be described as follow:
\begin{equation}
	a^{\dagger} \left | \varnothing_{n} \right>_{in} \rightarrow \sum_{i=1}^{4} g_{ni} a^{\dagger} \left | \varnothing_{i} \right>_{o}
\end{equation}
where the left hand side denotes the creation of a photon in n$th$ input waveguide and the right hand side describes how this initial state evolves in space and appears in i$th$ output waveguide. The parameter $ g_{ni} $ denotes element $ (n,i) $ of the quantum transfer matrix, $ \textbf{G} $.

Figure \ref{fig1}(e) illustrates a photon created in the 1$st$ input waveguide of the C-NOT gate. The injected photon evolves as follow
\begin{equation}
	\psi^w_{in} =a^{\dagger} \left | \varnothing_{1} \right>_{in} \rightarrow  \psi^w_{out} = g_{11}a^{\dagger} \left | \varnothing_{1} \right>_{o} + g_{12}a^{\dagger} \left | \varnothing_{2} \right>_{o} + g_{13}a^{\dagger} \left | \varnothing_{3} \right>_{o} + g_{14}a^{\dagger} \left | \varnothing_{4} \right>_{o}
\end{equation}
or
\begin{align}
	\left[  \begin{matrix}
		a^{\dagger} \left | \varnothing_{1} \right>_{out} \\
		0 \\
		0\\
		0
	\end{matrix} \right]	\rightarrow   \left[  \begin{matrix}
		g_{11} & g_{12} & g_{13} & g_{14} \\
		0 & 0 & 0 & 0 \\
		0 & 0 & 0 & 0\\
		0 & 0 & 0 & 0
	\end{matrix} \right]  \left[  \begin{matrix}
		a^{\dagger} \left | \varnothing_{1} \right>_{o} \\
		a^{\dagger} \left | \varnothing_{2} \right>_{o} \\
		a^{\dagger} \left | \varnothing_{3} \right>_{o}\\
		a^{\dagger} \left | \varnothing_{4} \right>_{o}
	\end{matrix} \right]
\end{align}
where in an ideal lossless case $ \sum_{i=1}^{4} \left | g_{ni}  \right |^2 = 1,~ n:1\rightarrow 4 $.

Next, we assume two photons are created at input ports $l$ and $k$ where $ l,~k = 1,~2,~3$ or $ 4 $. The input state can be described as $\psi^w_{in}  = a^{\dagger} \left | \varnothing_{l} \right>_{in} \times a^{\dagger} \left | \varnothing_{k} \right>_{in} $, accordingly, the two photons randomly walk and evolve to the following state at the gate's output:
\begin{align}
	\psi^w_{in}   \rightarrow	 \psi^w_{out} = \left[\sum_{i=1}^{4} g_{li} a^{\dagger} \left | \varnothing_{i} \right>_{o}  \right] \times \left[\sum_{j=1}^{4} g_{kj} a^{\dagger} \left | \varnothing_{j} \right>_{o} \right]  
\end{align}

The probability of detecting two photons simultaneously at output waveguides $m$ and $n$ can be described by the following twofold probability expression
\begin{equation}
	\Gamma^w_{mn} = 	\left < \psi^w_{out} \right | a^{\dagger}_m a^{\dagger}_n a_m a_n 	\left | \psi^w_{out} \right >
\end{equation}
where $ a^{\dagger}_{m/n},~a_{m/n} $ are the creation and annihilation operators at waveguides $m$ and $n$. The twofold probability demonstrates the correlation between simultaneous photons appearance at output waveguides $m$ and $n$. This detection scheme is based on the joint probability of two photons detection at the same time \cite{Gard}. The twofold probability at out output wavegides $m$ and $n$ is symmetrically equal to the twofold probability at the output waveguides $n$ and $m$, $\Gamma^w_{mn} = \Gamma^w_{nm}$. It is important to stress that, mathematically the twofold probability for the proposed C-NOT gate is a $4\times 4$ matrix. Assuming injected photons at port $l$ and $k$, the corresponding correlation matrix can be obtained as follow:
\begin{equation}
	\Gamma^{\mathrm{w}}= \left[  \begin{matrix}
		\left|  g_{l1} g_{k1} \right |^2 &   \left|  g_{l1} g_{k2} + g_{l2} g_{k1} \right |^2  &  \left|  g_{l1} g_{k3} + g_{l3} g_{k1} \right |^2  &   \left|  g_{l1} g_{k4} + g_{l4} g_{k1} \right |^2\\
		\left|  g_{l1} g_{k2} + g_{l2} g_{k1} \right |^2 &  \left|  g_{l2} g_{k2} \right | &  \left|  g_{l2} g_{k3} + g_{l3} g_{k2} \right |^2 &  \left|  g_{l2} g_{k4} + g_{l4} g_{k2} \right |^2 \\
		\left|  g_{l1} g_{k3} + g_{l3} g_{k1} \right |^2 & \left|  g_{l2} g_{k3} + g_{l3} g_{k2} \right |^2 &
		\left|  g_{l3} g_{k3}  \right |^2 &
		\left|  g_{l3} g_{k4} + g_{l4} g_{k3} \right |^2 \\ 
		\left|  g_{l1} g_{k4} + g_{l4} g_{k1} \right |^2 &
		\left|  g_{l2} g_{k4} + g_{l4} g_{k2} \right |^2 &
		\left|  g_{l3} g_{k4} + g_{l4} g_{k3} \right |^2 &
		\left|  g_{l4} g_{k4} \right |^2
	\end{matrix} \right]	
\end{equation}

In the above correlation matrix, each element $ \Gamma^w_{m,n} $ represents the probability of detecting exactly one photon at waveguide $m$ and one photon at waveguide $n$ at the output of the gate. This occurs when two photons are simultaneously injected into the input waveguides $l$ and $k$.

Next, we propose a hybrid photonic environment consisting of an optimized distribution of silicon dioxide particles in a silicon host. This photonic structure serves as a bosonic bath to demonstrate a medium for photons' continuous time random walk. Then we use the above formalism to calculate its correlation matrix and quantify gate's success rate.


\subsubsection{C-NOT Gate Photonic Realization}

Different photonic platforms have been proposed for the development of quantum gates. It is worth mentioning that in Ref. \cite{Ralph}, a probabilistic C-NOT gate for photons was proposed using a beamsplitter architecture, with a success probability of 1/9. Similarly, in Ref. \cite{D18}, a quantum photonic C-NOT gate was proposed via parallel waveguides, also with a success probability of 1/9. Recently, we introduced photonic logic gates using adjoint-method-assisted inverse design optimization for classical photonic computing \cite{Hassani_JLT}. These gates are composed of linear optical materials such as silicon and silicon dioxide, and enable fundamental logic operations such as the X-gate, AND-gate, OR-gate, as well as more complex operations like the C-NOT gate.

In the design of a gate, the optimization of scatterer particle distribution within the host medium is crucial. This can be achieved by utilizing basis spline functions and the Cox-de Boor recursion formula \cite{Boor,Cox}:
\begin{align}
	&N^{i,0}(u) =
	\begin{cases}
		1 & u_i < u<u_{i+1} \\
		0 & \text{otherwise }
	\end{cases} \nonumber \\&
	N^{i,k}(u) = \frac{u-u_i}{ u_{i+k} -u_i } N^{i,k-1}(u) + \frac{u_{i+k+1}-u }{ u_{i+k} -u_{i+1} } N^{i+1,k-1}(u)
\end{align}
where $ N^{i,k_x} (x) $ is a B-spline of degree $k$ and $ u_0,~u_1,\cdots, ~u_m  $ are the knots that defines the segments where b-spline functions are defined. The gate is implemented in a 2D setting, where a surface function $\Psi(x, y)$ is defined using B-spline functions. The level-set method is employed for topology optimization, where $\Psi(x, y)$ defines the area of the gate under optimization such that 
\begin{equation}
	\Psi(x, y) = \sum_{i=0}^{n_x} \sum_{j=0}^{n_y} N^{i,k_x} (x) N^{j,k_y}(y) P_{ij}
\end{equation}
where  $N^{i,k_x} (x)$ and $ N^{j,k_y}(y)$ are the B-spline functions of degree $k_x$ and $k_y$ and $ P_{ij} $ determines the B-spline functions wights in the above series \cite{Hassani_JLT,Zongfu2}. The main body of the gate is shaped based on the permittivity distribution described by the conditional equation:
\begin{align}\label{eps_xy}
	\epsilon(x,y )=
	\begin{cases}
		\epsilon_{SiO_2} & \Psi(x,y) < 0 \\
		\epsilon_{Si} & \Psi(x,y) > 0
	\end{cases} 
\end{align}

Here, $\epsilon_{Si}$ and $\epsilon_{SiO_2}$ represent the permittivity of silicon (host) and silicon dioxide (scatterers), respectively. The Maxwell's equation in the frequency domain is then solved to obtain the field distribution:
\begin{equation}
	\boldsymbol{\nabla} \times \boldsymbol{\nabla} \times \textbf{E}   - \omega^2 \epsilon(x,y) \mu_0 \textbf{E} = -i\omega \mu_0 \textbf{J}
\end{equation}
where $\textbf{J}$ represents the current source density that represents the spatial profile of the input light.

The optimization process involves initiating a random distribution of silicon dioxide within the silicon host medium. Maxwell's equations are then solved with an appropriate current distribution along the input waveguide pair, representing a specific input state. The resulting electric field distribution, referred to as the "forward field" ($\textbf{e}$), is compared against the target performance (gate's fidelity in this work), and the cost function $\mathcal{C}$ is evaluated. The goal of optimization is to minimize $\mathcal{C}$ by adjusting the silicon dioxide distribution withing the silicon host. The gradient of the cost function with respect to this parameter is calculated using the following equation \cite{Zongfu2}:

\begin{align}
	\frac{\partial \mathcal{C}}{ \partial P_{ij} } =  \iint_{x,y}  \frac{\partial \mathcal{C}}{ \partial \epsilon(x,y) } \delta(\Psi(x,y))   N^{i,k_x} (x) N^{j,k_y}(y) dxdy
\end{align}
where $\delta(\Psi(x,y))$ is a Dirac delta function that is non-zero only where $\Psi(x,y)$ passes zero (i.e., the boundaries between silicon and silicon dioxide). The first term in the equation represents the geometrical gradient of the cost function, which can be calculated using the adjoint method.

Following the determination of the cost function, the adjoint equation $\textbf{A}^T  \boldsymbol{\lambda} = -\left(\partial_{e} \mathcal{C}\right)^T$ is solved to obtain the "backward field" ($\boldsymbol{\lambda}$). This allows the error term $\partial_{e} \mathcal{C}$ to propagate backwards from the output side to the input side. With both the forward and backward fields, the structural gradient $\partial_{\epsilon(x,y)} \mathcal{C}$ is computed as $\partial_{\epsilon(x,y)} \mathcal{C} = 2 Re \left\{    \boldsymbol{\lambda} \odot \textbf{e} \right\} $, \cite{Hassani_JLT,Zongfu2,Zongfu3}. The medium under optimization is then updated using B-spline functions, and the level-set method is employed to evolve the boundaries between silicon and silicon dioxide in a manner that improves the device's performance. We provided more details on this method in designing photonic gates in \cite{Hassani_JLT}. In the optimization process, the cost function is evaluated by the equation:
\begin{equation}
	\mathcal{C} = \iint  \left | P_m - P_t  \right | /P_t  dl
\end{equation}
where $ P_m$ and $P_t$ are the measured and target power distribution in the output photonic bits in each iteration. The integral in the above equation is across the output waveguide. This quantity $ \iint \left | P_m - P_t  \right | dA $ is sometimes known as Kolmogorov distance \cite{D14}.

\begin{figure}[ht!]
	\centering\includegraphics[width=11cm]{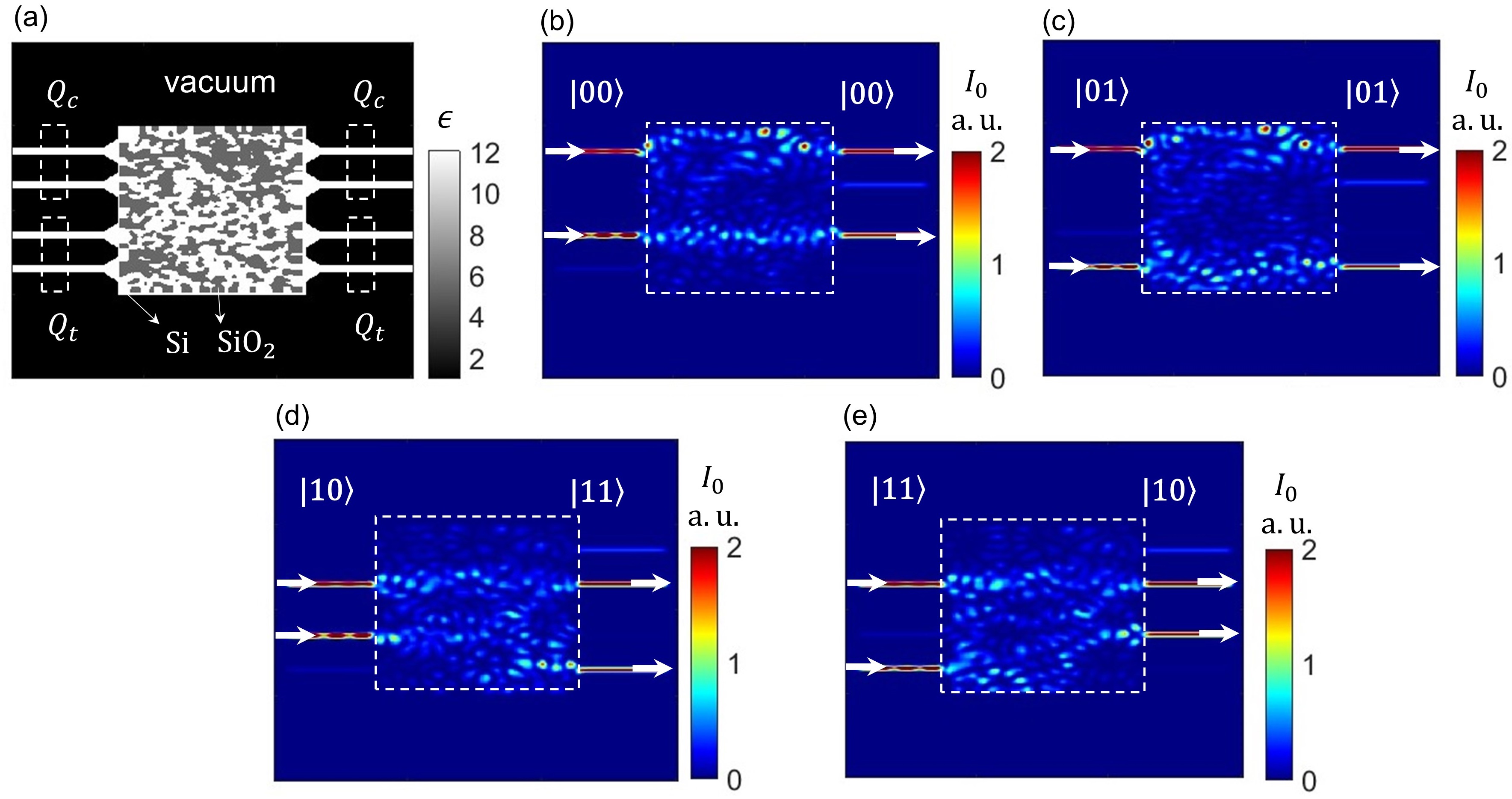}
	\caption{(a) Proposed C-NOT gate with two input/output control/target qubits indicated by the dashed lines. This gate consists of a host silicon medium and a group of distributed SiO2 scatterers. (b), (c), (d) and (e) Shows the electric field intensity distribution for the principle logic operations of the proposed C-NOT gate $  \left| 00 \right > \rightarrow   \left| 00 \right > $, $  \left| 01 \right > \rightarrow   \left| 01 \right > $,  $  \left| 11 \right > \rightarrow   \left| 10 \right > $ and  $  \left| 10 \right > \rightarrow   \left| 11 \right > $, respectively. The white dashed boxes in these panels determines the gate's main body where the two created photons randomly walk.}\label{fig2}
\end{figure}

Here, we designed a C-NOT gate and examine its efficiency by calculating the quantum correlation matrix using the above-described method. Figure \ref{fig2}(a) illustrates the proposed C-NOT gate. The input and output qubits are connected through the main body of the gate comprising silicon and silicon dioxide. The physical dimensions of the gate's main body are $2.5\lambda \times 3.2\lambda$ where $ \lambda = 1.55\mu m $, is the operational wavelength. This gate is optimized to receive photons with TM (in-pane) polarization. The optimization of the silicon dioxide distribution inside the silicon host has been performed based on the gate's fidelity. Figure \ref{fig2}(b), (c), (d) and (e) demonstrate the electric field intensity distribution for different C-NOT gate logic operations. In each panel, the corresponding evolution is indicated as an inset.

In Fig. \ref{fig2}(b), the field intensity distribution is demonstrated when the input target and control qubits are prepared in the state $ \left| Q_c Q_t \right > = \left| 00 \right > $, meaning two photons are created at the upper waveguides of the two input pairs, $ \psi^w_{in} = a^{\dagger} \left| \varnothing_{1} \right>_{in} \times a^{\dagger} \left| \varnothing_{3} \right>_{in} $. The two photons randomly walks and interfere through the main body of the gate and appear at 1$st$ and 3$rd$ output waveguides, indicating the output state $ \left| Q_c Q_t \right > = \left| 00 \right > $. The other panels in Fig. \ref{fig2} display the same result but for the other principal logic operations.

\begin{figure}[ht!]
	\centering\includegraphics[width=14cm]{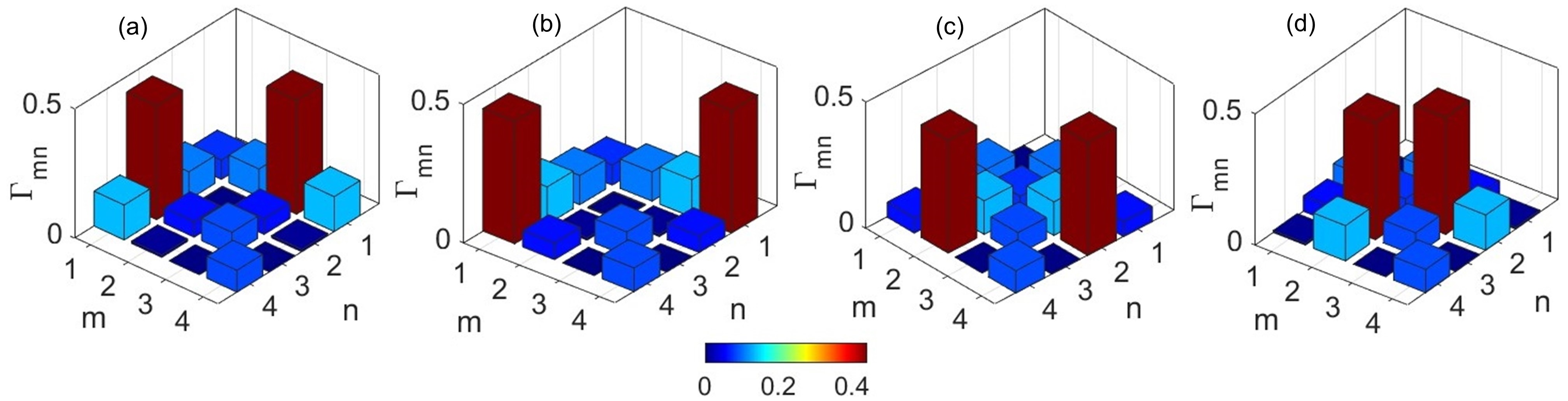}
	\caption{Quantum correlation matrix elements of the proposed C-NOT gate for all principal operations. Each element $ \Gamma_{m,n} $ represents the probability of detecting exactly one photon at waveguide $m$ and one photon at waveguide $n$ simultaneously at the gate's output, when the input state is (a)- $ \left | 00 \right>  $, (b)- $ \left | 01 \right>  $, (c)- $ \left | 10 \right> $ and (d)- $ \left | 11 \right>  $.   }\label{fig3}
\end{figure}

Eventually Fig. \ref{fig3} illustrates the quantum correlation matrix elements of the proposed C-NOT gate for all principal operations. Each element $ \Gamma^w_{m,n} $ represents the probability of detecting exactly one photon at waveguide $m$ and one photon at waveguide $n$ at the gate's output, when the input state is (a) $ \left | 00 \right>  $, (b) $ \left | 01 \right>  $, (c) $ \left | 10 \right> $ and (d) $ \left | 11 \right>  $. For instance, the output state corresponding to the input state $ \left| 00 \right> $ is $ \left| 00 \right> $, meaning the output photons must appear in waveguides 1 and 3. In terms of the correlation matrix, it is the elements $ \Gamma_{13}= \Gamma_{31} $ that would take the largest values. A visual inspection of Fig. \ref{fig3}(a) reveals that columns related to $ m, n=1,3 $ have heights (success probability) approximately equal to $ \approx 0.46 $. This is also true for other input states. As another example the output state corresponding to the input state $ \left| 11 \right> $ must be $ \left| 10 \right> $, meaning the output photons must appear in waveguides 2 and 3, and inspection of Fig. \ref{fig3}(d) also reveals that columns related to $ m, n=2,3 $ show success probability approximately equal to $ \approx 0.46 $. To sum up this section, it has been shown that the proposed C-NOT gate has a success probability of $ \approx 0.46$ ($\approx 4.14\times $ larger than what reported in \cite{Ralph} and \cite{D18}).

\begin{figure}[ht!]
	\centering\includegraphics[width=6.7cm]{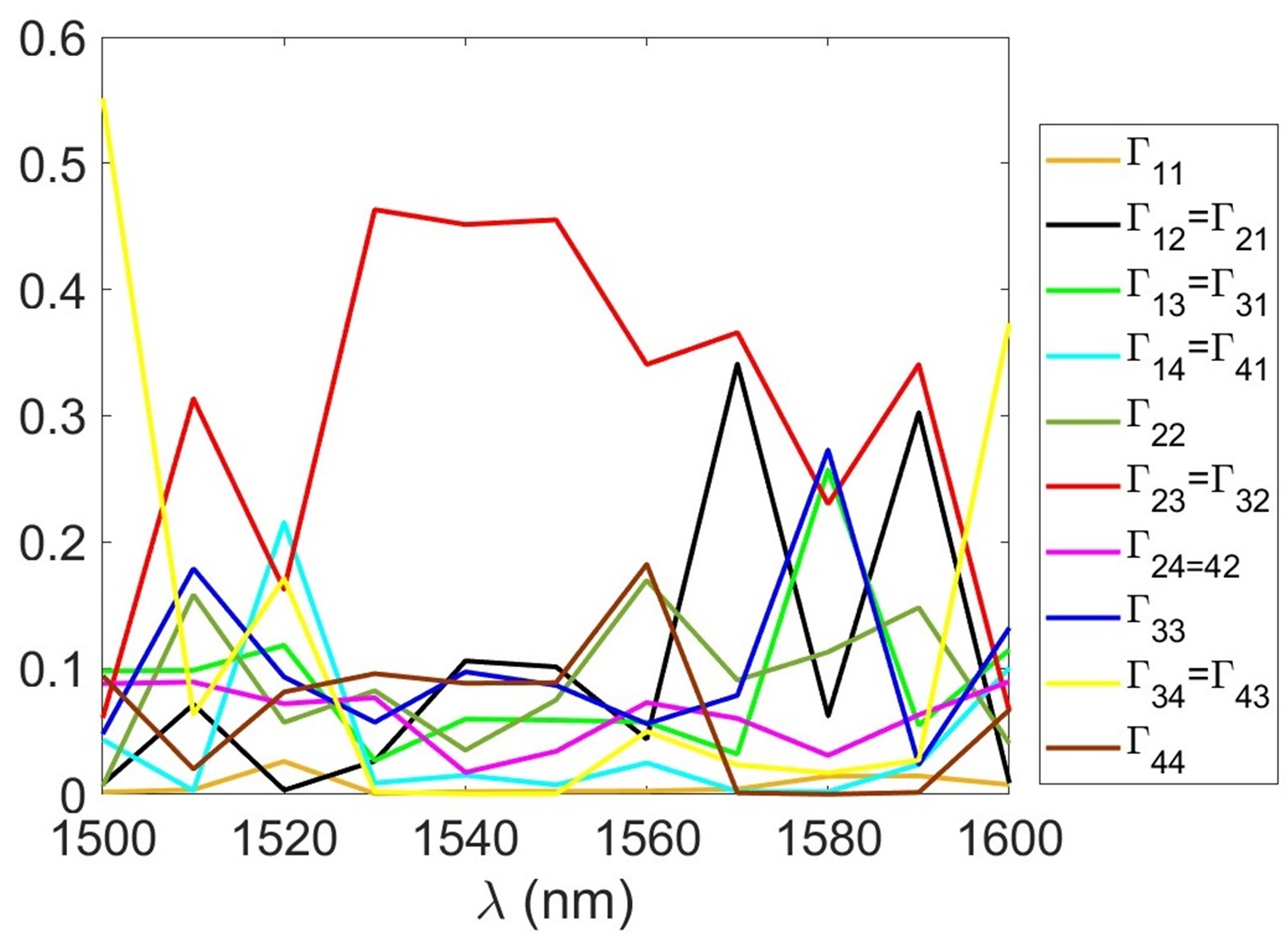}
	\caption{Correlation matrix elements of the C-NOT gate in Fig. \ref{fig2} versus wavelength for the principle logic operation  $ \left | 11 \right > \rightarrow \left | 10 \right >$. }\label{fig3p}
\end{figure}

Next, we investigate the frequency response of this C-NOT gate by calculating the correlation matrix elements for one of its principal logic operations over a wide range of wavelengths, for instance $\left|11\right> \rightarrow \left|10\right>$. Figure \ref{fig3p} demonstrates the wide-band responses of this logic operation. The solid red line indicates the element $\Gamma_{23} = \Gamma_{32}$, indicating the probability of detecting one photon at output waveguides 2 and 3 simultaneously. Compared to Fig. \ref{fig3}(d) which shows the result at the design wavelength $\lambda = 1550~nm$, the gate still performs well with $\Gamma_{23} = \Gamma_{32} \approx 0.46$, over a wide range around the target wavelength.

\subsubsection{C-NOT Gate Fidelity}

Due to the quantum nature of a multi-photon system, it is expected that photonic logic gates would have different outcomes with different probabilities which can be described by quantum correlation matrix. For instance, inspection the quantum correlation matrix elements in Fig. \ref{fig3} reveals that there are unwanted coincidences as well. These unwanted events occur due to the leakage of photons into the unwanted waveguides in the output of the C-NOT gate shown in Fig. \ref{fig2}. Therefore, it is necessary to evaluate the fidelity of the proposed C-NOT gate. The gate fidelity is defined as the Hilbert-Schmidt inner product between the target unitary gate operator $U_0$ and the unitary operator $U$ of the designed C-NOT gate. The gate fidelity can be calculated using the formula \cite{D18}:

\begin{equation}\label{fidelity}
	F(U_0, U) = \frac{{\mathrm{Tr}(U_0^{\dagger} U)}}{N}
\end{equation}
here, $N$ represents the dimension of the logical space, which is 4 for a two-qubit C-NOT gate. 

To calculate the fidelity, we need to obtain the system's Hamiltonian. In Fig. \ref{fig2}(b), if the input state is $ \left |00  \right> $, the goal is to receive the photons in waveguides 1 and 3, resulting in an output state of $ \left |00  \right> $. However, if photons are detected in the undesired waveguides, it means that although the desired state $ \left | 00 \right> $ is detected, there is a probability for the presence of other states as well. Therefore, the output state can be described as $  c^{00}_{00} \left | 00 \right> + c^{00}_{01} \left | 01 \right> + c^{00}_{10} \left | 10 \right> + c^{00}_{11} \left | 11 \right>$, where $ c^{lk}_{ij} $ is the expansion coefficient of the output eigen-state such that the superscript indicates the input state and the subscript denotes the corresponding eigen-state outcome. The same is true for other input cases. Based on this we construct the Hamiltonian as follow:

\begin{align}
	H_{CNOT} &= \left[ c^{00}_{00} \left | 00 \right> + c^{00}_{01} \left | 01 \right> + c^{00}_{10} \left | 10 \right> + c^{00}_{11} \left | 11 \right> \right] \left<  00 \right| \nonumber \\& 
	+ \left[ c^{01}_{00} \left | 00 \right> + c^{01}_{01} \left | 01 \right> + c^{01}_{10} \left | 10 \right> + c^{01}_{11} \left | 11 \right> \right] \left<  01 \right| \nonumber \\&
	+ \left[ c^{10}_{00} \left | 00 \right> + c^{10}_{01} \left | 01 \right> + c^{10}_{10} \left | 10 \right> + c^{10}_{11} \left | 11 \right> \right] \left<  10 \right| \nonumber \\&
	+ \left[ c^{11}_{00} \left | 00 \right> + c^{11}_{01} \left | 01 \right> + c^{11}_{10} \left | 10 \right> + c^{11}_{11} \left | 11 \right> \right] \left<  11 \right|
\end{align}
it can be confirmed that in case of no photon detection in unwanted waveguides all $ c^{lk}_{ij} $ are zero except $ c^{00}_{00} = c^{01}_{01} = c^{10}_{11} = c^{11}_{10} = 1 $, thus above matrix reduces to the known ideal C-NOT gate Hamiltonian, $ H_0 = [1~0~0~0;0~1~0~0; 0~0~0~1; 0~0~1~0] $. As an example the Hamiltonian of the proposed optimized C-NOT gate in Fig. \ref{fig2} is (to two decimal places):

\begin{align}
	H_{CNOT} &= \left[  \begin{matrix} 
		0.85 &  0 &  0   & 0.13\\
		0.01 & 0.85 & 0.13 & 0  \\
		0.13 & 0 & 0 & 0.85 \\ 
		0 & 0.13 & 0.85 & 0.01 
	\end{matrix} \right]
\end{align}
using the provided Hamiltonian and the C-NOT gate target Hamiltonian, it can be shown that the fidelity of the proposed C-NOT gate is approximately $F(U_0, U) \approx 0.95$.


\subsection{Single Random Walker: X-Gate}

\begin{figure}[ht!]
	\centering\includegraphics[width=11cm]{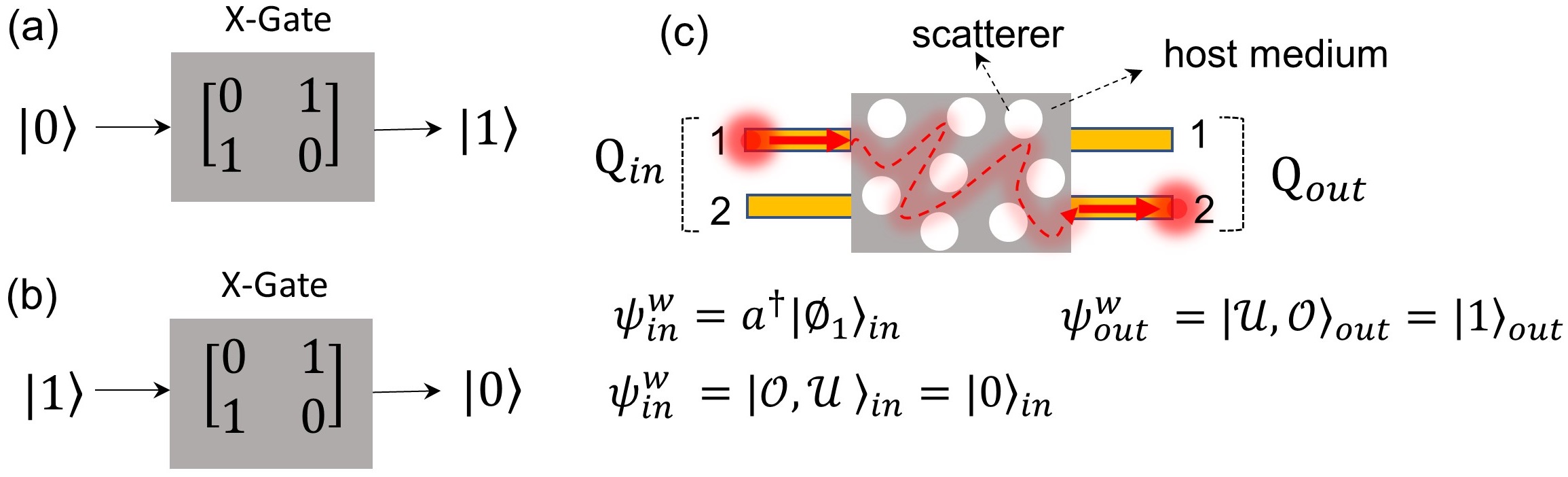}
	\caption{(a)-(b) Computational and (c) dual-rail physical representations of the X-gate. The main operation of the X-gate is to flip the input state. As shown in (c), the X-gate accomplishes this by swapping the positions of the input photons in the output through a two-dimensional photon random walk. }\label{fig4}
\end{figure}

The computational representation of an X-gate is depicted in Fig. \ref{fig4}(a) and (b). This gate, featuring one input qubit and one output qubit, operates by flipping the input state. Figure \ref{fig4}(c) depicts the dual-rail representation, where two sets of parallel waveguides are used as the input and output qubits, $Q_{in}$ and $ Q_{out} $, respectively. The main body of the gate connects the input and output qubits and acts as the environment for the input photon (represented by the red dot) to undergo a two-dimensional random walk. Since each input qubit corresponds to one photon, this gate utilizes a single random walker to preform a quantum logic.

\subsubsection{Correlation Functions: One-fold Probability}

Similar to the C-NOT gate and multiple walker, the input state can be represented by the creation of a photon at a specific waveguide of the input waveguide pair. If the photon is created in the top waveguide, the input state is $\psi^w_{in} = a^{\dagger} \left| \varnothing_{1} \right>_{in} =  \left | \mathcal{O}, \mathcal{U} \right >$, which is recognized as $\left| 0 \right>_{in}$. Otherwise, if the photon is created in the bottom waveguide, the input state is $\psi^w_{in} = a^{\dagger} \left | \varnothing_{2} \right>_{in} = \left | \mathcal{U}, \mathcal{O} \right >$, which is recognized as $\left| 1 \right>_{in}$.

For instance in Fig. \ref{fig4}(c) the input state is prepared at $\left| 0 \right>$ by creating a photon in the upper waveguide, and this photon appears in the output bottom waveguide, indicating output state as $\left| 1 \right>$. Thus, the probability of detecting one photons at a specific output waveguides $n$ can be described by the following one-fold probability expression
\begin{equation}
	\Gamma^w_{n} = 	\left < \psi^w_{out} \right |  a^{\dagger}_n  a_n 	\left | \psi^w_{out} \right >
\end{equation}
where $ a^{\dagger}_{n},~a_{n} $ are the creation and annihilation operators at waveguide $n$. In single walker case, the mean photon number is equal to the one-fold probability that waveguide $n$ receives a photon. The probability distribution over all detectors provides all the information.

The created photon in the input waveguide $i$, randomly walks through the host-scatterers and will appear in the output waveguides. Therefore its evolution can be described as follow:
\begin{equation}
	\psi^w_{in} =a^{\dagger} \left| \varnothing_{i} \right>_{in} \rightarrow  \psi^w_{out} = \sum_{n=1}^{2} g_{in}a^{\dagger} \left| \varnothing_{n} \right>_{o}
\end{equation}
therefore depending on the input photon site, waveguide i$th$, the correlation function becomes a one dimensional vector $ \Gamma^w = \left[ \left | g_{i1} \right |^2 ~ \left | g_{i2} \right |^2 \right]  $, where the first subscript $ i = 1,~2 $ is the input waveguide index and the second subscript indicate the output waveguide index. Next we apply this formalism on a single random walker to perform quantum logic in an X-gate.

\subsubsection{X-Gate Photonic Realization}

\begin{figure}[ht!]
	\centering\includegraphics[width=11cm]{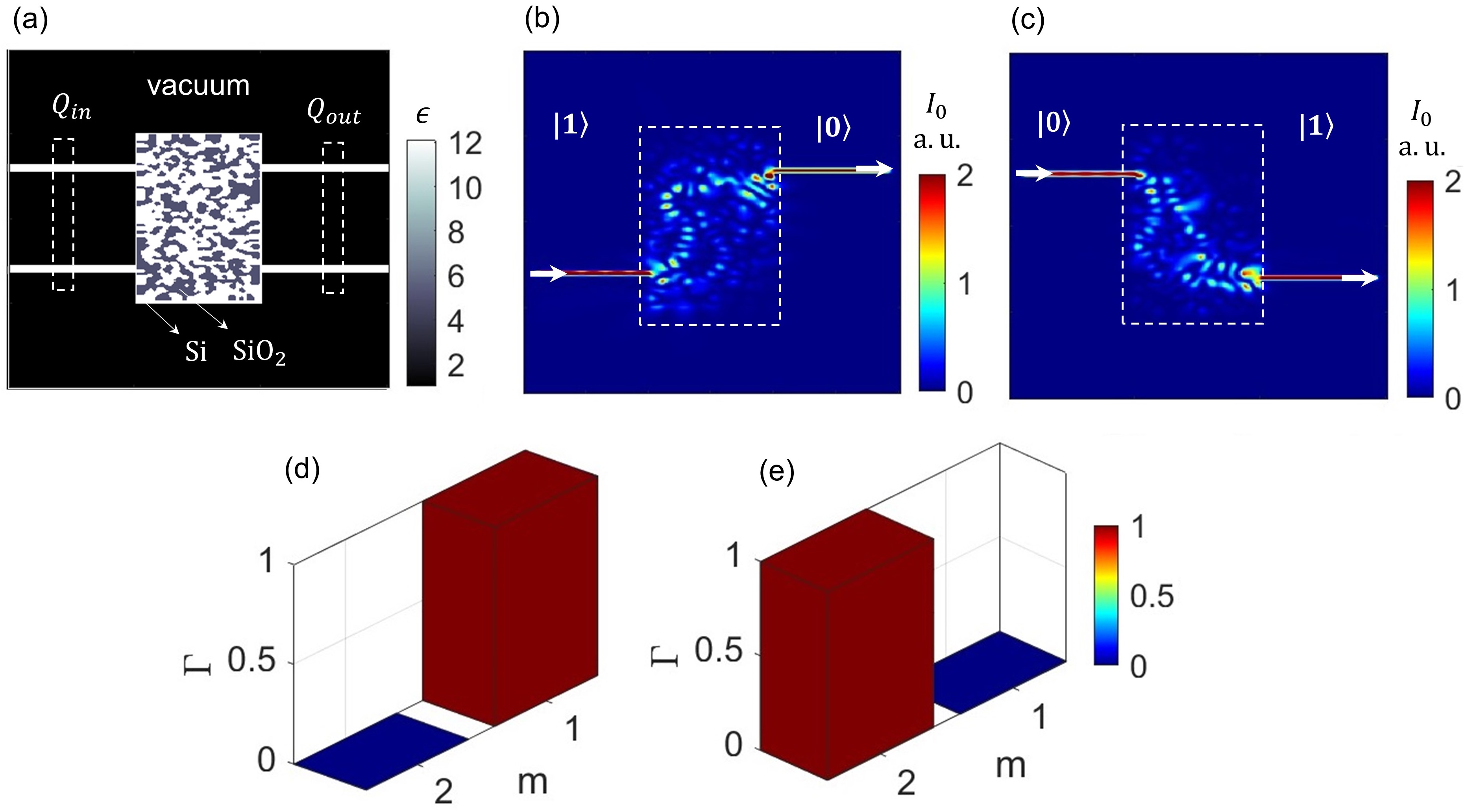}
	\caption{(a) Proposed X-gate with one input and output qubits indicated by the dashed lines. Similar to the C-NOT gate, this gate also consists of a host silicon medium and a group of distributed SiO2 scatterers. (b) and (c) Shows the electric field intensity distribution for the principle logic operations of the proposed X-gate $  \left| 1 \right > \rightarrow   \left| 0 \right > $ and $  \left| 0 \right > \rightarrow   \left| 1 \right > $, respectively. The white dashed boxes in these panels determines the gate's main body where the input photon experiences a two-dimensional random walk. (d) and (e) Illustrates the correlation function for the two principal operations. Each element $ \Gamma_{m} $ represents the probability of detecting exactly one photon at the output waveguide $m$, when with input state is (d)- $ \left | 1 \right>  $, (e)- $ \left | 0 \right>  $.}\label{fig5}
\end{figure}

Figure \ref{fig5}(a) illustrates the physical design of an X-gate. A pair of single mode silicon waveguides is utilized as the input and output qubits. Similar to the C-NOT gate, this gate is also primarily composed of a hybrid material combination of silicon and an optimized distribution of silicon dioxide, see our recent work on the optimization process \cite{Hassani_JLT} The total in-plane physical size is approximately $ 1.7\lambda \times 3.2\lambda $ with $ \lambda = 1.55~ \mu m $ and the polarization of the input waveguide is TM. 

Figure \ref{fig5}(b) displays the field intensity distribution when the input qubits is prepared in the state $ \left| Q_{in} \right > = \left| 1 \right > $, meaning one photon is launched into the bottom waveguides of the input pair, $ \psi^w_{in} = a^{\dagger} \left| \varnothing_{2} \right>_{in} $. This photon randomly walks through the main body of the gate and appears at 1$st$ output waveguide, indicating the output state $ \psi^{qb}_{out} = \left| Q_{out}\right > = \left| 0 \right > $. The other panels in Fig. \ref{fig5}(c) illustrates the alternative evolution of the X-gate, wherein the input qubit is in the $ \left | 1 \right > $ state. As can be seen, the output state is $ \left | 0 \right > $.

Figure \ref{fig5}(d) and (e) illustrate the correlation function elements of the proposed X-NOT gate. Each element, denoted as $\Gamma^w_{m}$, represents the probability of detecting exactly one photon at waveguide $m$ at the gate's output, given the input state is (d) $\left | 1 \right>$ and (e) $\left | 0 \right>$. For the input state $\left | 1 \right>$, the output state corresponds to $\left | 0 \right>$, indicating that the output photon must appear in waveguide 1. In terms of the correlation function elements, the first element has the largest value. Panel (d) clearly shows that the first column is $\approx 1$, while the other column is $\approx 0$, indicating a success probability of $\approx 1$ in receiving the input photon in the expected output waveguide. Similarly, for the input state $\left | 0 \right>$, the output photon must appear in the output waveguide 2. By inspecting Figure \ref{fig5}(e), it is evident that the second column is $\approx 1$, while the other column is $\approx 0$. Therefore, the success probability for the X-NOT gate operation is also $\approx 1$.

\begin{figure}[ht!]
	\centering\includegraphics[width=10cm]{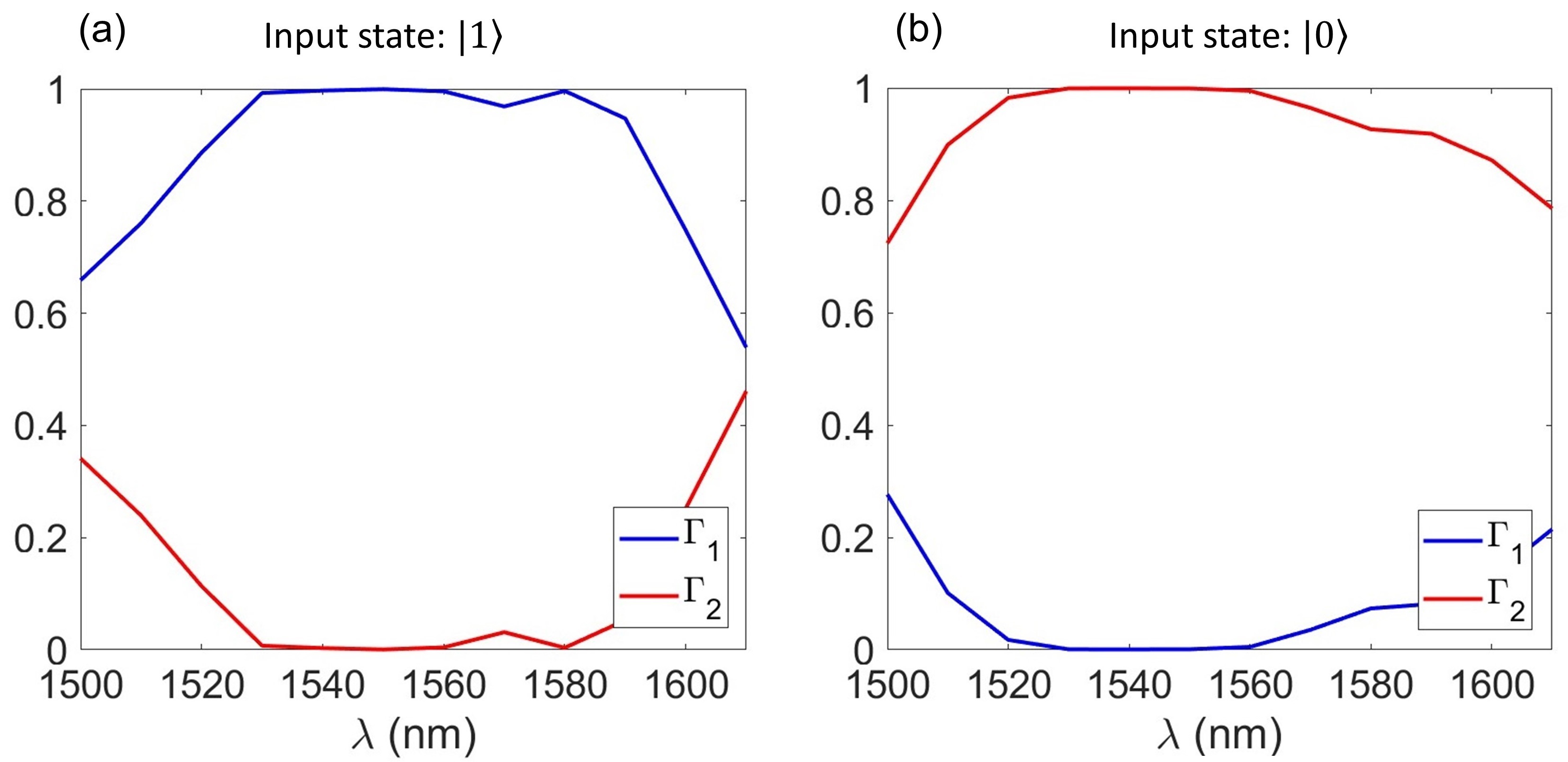}
	\caption{Correlation function of the X-gate in Fig. \ref{fig5} versus wavelength when the input state is set to be (a) $\left|1\right>$ and (b) $\left|0\right>$.  }\label{fig5p}
\end{figure}

We also show the frequency response of the X-gate correlation function. The correlation function versus wavelength is calculated in Figure \ref{fig5p}, when the input state is set to be $\left|1\right>$ and $\left|0\right>$. As is clear, although the design wavelength is $\lambda = 1550~nm$, the element $\Gamma_{1}$ for the input state $\left|1\right>$ and $\Gamma_{2}$ for the input state $\left|0\right>$ remain larger than 0.9 over a wide range of wavelengths, depicting the possible wide-band operation of the X-gate. Next, we calculate the X-gate fidelity.

\subsubsection{X-Gate Fidelity}

In Fig. \ref{fig5}(b), when the input state is $ \left| 1 \right> $, the goal is to receive the photon in the top output waveguide, resulting in an output state of $ \left|0 \right> $. However, if there is a small amount of leakage in the bottom waveguide that is detected, it means that in addition to the desired state $ \left|0 \right> $, its flipped version $ \left|1 \right> $ might also be detected. Therefore, the output state can be described as $ c^{1}_{0}  \left|0 \right> + c^{1}_{1} \left|1 \right>   $, where $c^{i}_{j}$ is the expansion coefficient of the output state with superscript indicating the input state and the subscript denoting the corresponding output eigen-state. Based on this, the proposed X-gate Hamiltonian can be constructed as follows:

\begin{equation}
	H_{X} = \left[ c^{0}_{0}  \left|0 \right> + c^{0}_{1} \left|1 \right> \right] \left< 0 \right | + \left[ c^{1}_{0}  \left|0 \right> + c^{1}_{1} \left|1 \right> \right] \left< 1 \right |
 \end{equation}

Note that the above Hamiltonian reduces to the known ideal X-gate Hamiltonian if there is no leakage into the unwanted waveguide. In this case all $ c^{i}_{j} $ are zero except $ c^{0}_{1} = c^{1}_{0} = 1 $, leading to the ideal X-gate Hamiltonian $ H_0 =   \left|1 \right>  \left< 0 \right |+  \left|0 \right>  \left< 1 \right | $. By following the same procedure presented in Sec. 2.1.3, it can be shown that the proposed X-gate in Fig. \ref{fig5} can achieve $F(U_0, U) \approx 1$.


\section{Geometrical Imperfection Effects on Correlation Function}

\begin{figure}[ht!]
	\centering\includegraphics[width=13cm]{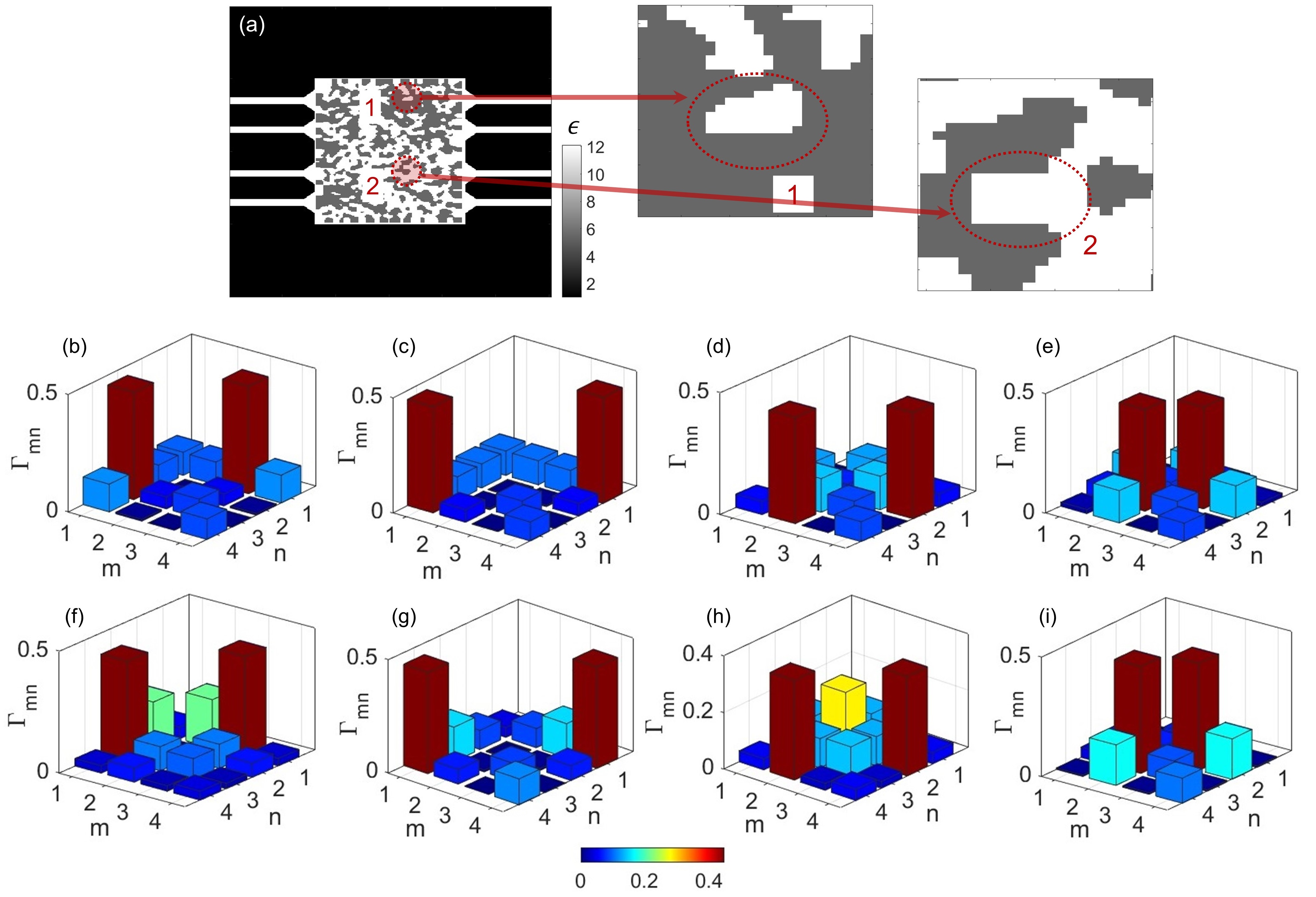}
	\caption{(a) The geometrical representation of the optimized C-NOT gate with its functionality shown is in Fig. \ref{fig2}. Two islands, with a relative size of $0.25 \mu m \times 0.25 \mu m$, are designated to be removed from the structure. These geometrical features are located in positions with higher field intensity, as seen in the field distribution for different logic operations in Figure \ref{fig2}. The middle row shows the quantum correlation matrix elements, $\Gamma_{mn}$, when island 1 is removed with the input state: (b) $\left|00\right>$, (c) $\left|01\right>$, (d) $\left|10\right>$, and (e) $\left|11\right>$. The last row shows the same results but when the second island is removed.  }\label{fig8}
\end{figure}

As can be seen in Figs. \ref{fig2} and \ref{fig5}, the optimized logic gates have intricate structures. Therefore, it is important to understand the effects of these small features, particularly within the manufacturing tolerance, on the operation of the logic gates. First, we will focus on the C-NOT gate shown in Fig. \ref{fig2}(a). Two different geometrical features or islands are chosen, with sizes in the range of $0.25 \mu m \times 0.25 \mu m$, as shown on the right side of Fig. \ref{fig8}(a). The selection criteria ensure that each island is situated in a region with high field intensity, as depicted in Fig. \ref{fig2}, since these areas play a significant role in controlling the propagation and interference of photons along the gate's body. The islands selected in \ref{fig8}(a) are made of silicons and to remove them, they are replaced by silicon dioxide.

Panels (b), (c), (d), and (e) in Fig. \ref{fig8} depict the C-NOT quantum correlation matrix elements when the first island in Fig. \ref{fig8}(a) is removed. Comparing these matrix elements with those calculated for the original C-NOT gate in Fig. \ref{fig3} reveals that the columns related to $m, n=1,3$ still have heights approximately equal to $0.46$, similar to the original C-NOT gate. However, removing the second island shown in Fig. \ref{fig8}(b) has observable effects on the gate's performance. As can be seen in panels (f), (g), (h), and (i) in Fig. \ref{fig8}, by removing the second island, although the heights of the columns related to $m, n=1,3$ drop only slightly, the columns related to unwanted coincidences become higher. It must be noted that the fabrication tolerance of silicon nanophotonics is usually better than $0.1 \mu m$ \cite{SiPh}, which is smaller than the size of the defects we removed. Therefore, our analysis indicates that the impacts 
of the delicate structures in the photonics gates could be acceptable. However one may conclude that depending on the size and position of the imperfection in the gate's main body, imperfections may impact differently on the gate functionality.

\begin{figure}[ht!]
	\centering\includegraphics[width=12cm]{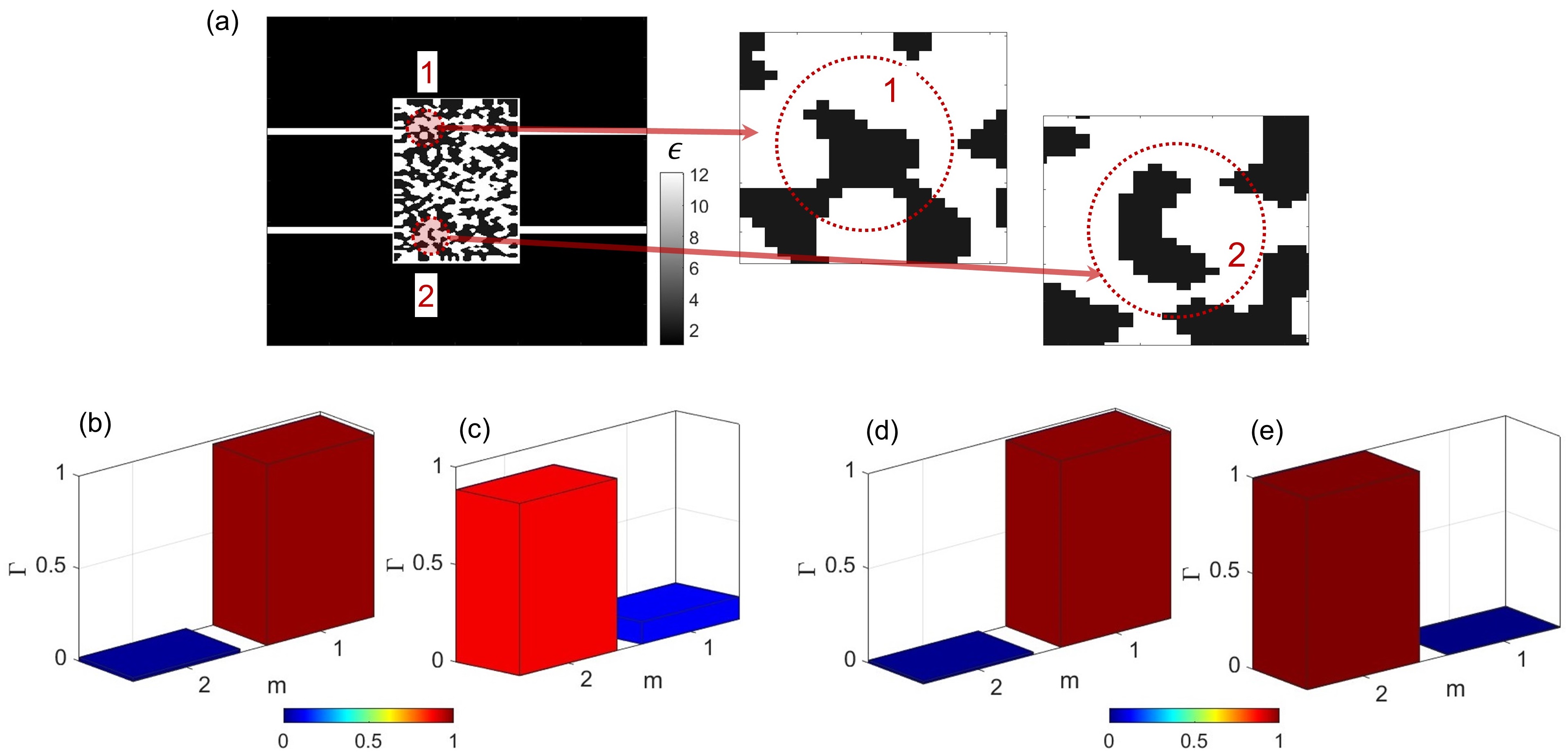}
	\caption{ (a) The geometrical representation of the optimized X-gate with its functionality is shown in Fig. \ref{fig5}. Two islands, with a relative size of $0.25\mu m \times 0.25\mu m$, are designated to be removed from the structure. (b)-(c) show the correlation function, when island 1 is removed with the input state $\left|1\right>$ and $\left|0\right>$, respectively. (d)-(e) show the same results but when the second island is removed.  }\label{fig9}
\end{figure}

Next, we conduct a similar analysis on the X-gate. Figure \ref{fig9}(a) illustrates the geometrical representation of this gate. In the field distribution for different logic operations shown in Figure \ref{fig5}, two islands are designated in areas where the field intensity is relatively high. The selected islands are made of silicon dioxide and to remove them, they are replaced by silicon.

In panels (b) and (c) of Figure \ref{fig9}, we present the correlation function when island 1 is removed, with input states $\left|1\right>$ and $\left|0\right>$, respectively. Comparing these correlation functions to those shown in Figure \ref{fig5}, we observe that only $ \Gamma_1 $ for input state $ \left| 0 \right> $ decreases from 1 to 0.88, and accordingly $ \Gamma_2 $ increases from 0 to 0.12. However, removing the second island does not result in any changes in the correlation function.

\section{Conclusion}

To sum up, this work demonstrates the feasibility of performing quantum logic operations using continuous-time two-dimensional random walkers within a hybrid photonic environment. We have proposed and analyzed the C-NOT gate, as a multi-qubit gate, and the X-gate, as a single-qubit gate, leveraging the dual-rail encoding scheme for the qubits. The gates are designed using the inverse design method, where photons randomly walk in a two-dimensional silicon host medium embedded with an optimized distribution of silicon dioxide scatterers. The detailed analysis of the quantum correlation function and gates fidelity underscores the reliability and effectiveness of these gates. It has been shown that the success rate of the C-NOT gate is approximately 0.46, with a fidelity around 0.95. Additionally, the success rate and fidelity of the X-gate are approximately 1 in a wide range of wavelengths around the design wavelength of $\lambda = 1550~nm$. This work is a step forward in the practical implementation of quantum computing using photonic platforms, with promising implications for future advancements in scalable and robust quantum computing architectures.



\end{document}